\documentstyle[multicol,epsfig,aps,prl,array]{revtex}
\begin{document}
\draft

\title{Universal Algebraic Relaxation of Velocity and Phase\\ 
in Pulled Fronts generating Periodic  or Chaotic States}

\author{Cornelis Storm$^1$, Willem Spruijt$^1$, Ute Ebert$^{1,2}$ 
  and Wim van Saarloos$^1$}
\address{$^1$Instituut--Lorentz, Universiteit Leiden, Postbus 9506,
  2300 RA Leiden, The Netherlands \\ 
  $^2$Centrum voor Wiskunde en Informatica,  Postbus 94079, 
  1090 GB Amsterdam, The Netherlands}
\date{\today} \maketitle

\begin{abstract}   
We investigate the asymptotic  relaxation of so-called pulled fronts
propagating 
into an unstable state. 
The ``leading edge representation'' of the equation of motion reveals 
the universal nature of their propagation mechanism and allows us to
generalize the universal algebraic velocity  
relaxation of uniformly translating fronts to fronts, that generate 
periodic or even chaotic states.
Such fronts in addition exhibit a universal algebraic phase relaxation.
We numerically verify our analytical predictions for the
Swift-Hohenberg and the Complex Ginzburg Landau equation.
\end{abstract}

\begin{multicols}{2}
  Many  systems, when driven sufficiently far from
  equilibrium, spontaneously organize themselves in coherent or
  incoherent patterns. The ubiquity of such structures in almost all
  fields of the natural sciences \cite{CH,bio} has inspired
  much of the recent scientific effort to uncover the various
  mechanisms underlying their behavior. Especially in physics, the
  insight from the seventies that in critical phenomena universality
  classes are determined essentially by the symmetry of the order
  parameter and the dimensionality, initially raised some hopes that
  there would be analogous broad universality classes in
  nonequilibrium pattern formation. 
  Over the last two decades, it has
  become clear, however, that such far-reaching universality does not
  exist: While there {\em are} various general dynamical and
  instability mechanisms, there is not always a sharp selection
  mechanism. Moreover, if there is sharp
selection, the particular mechanism may depend on the
  specific boundary conditions, initial conditions, etc.

 In the case of front propagation, there have, nevertheless,  
been several hints of a generic dynamical mechanism \cite{bj,vs2}: There is
  a large class of fronts propagating into an unstable state whose
  asymptotic velocity equals $v^*$, the asymptotic spreading
  velocity of linear perturbations about the unstable state. Such
  fronts  are called {\em pulled} fronts, 
  as they are ``pulled along'' by the leading edge of the profile
  whose dynamics is governed by the
  linearization about the unstable state
  \cite{stokes,paq,evs}. It is the purpose of this letter to show that
  within the subclass of pulled front propagation, a remarkable degree
  of universality does hold: Irrespective of whether such fronts are
  uniformly translating or generate periodic or chaotic patterns, the
  velocity $v(t)$ and phase  $\Gamma(t)$ of pulled fronts
  which emerge from steep initial conditions (falling off faster than
  $e^{-\lambda^* x} $ for $x\rightarrow \infty$), display a {\em
    universal} power law relaxation with time $t$, expressed by
\begin{eqnarray} \label{predv}
v(t)&  \equiv &  v^*+ \dot{X}(t) \\  \label{predv2}
\dot{X}(t)& = & -\frac{3}{2 \lambda^* t}
+\frac{3\sqrt \pi}{2\lambda^{*2} t^{3/2}}\mbox{Re}
\left(\frac{1}{ \sqrt{D} } \right) + {\cal O}\left(\frac{1}{t^{2}}\right)~, 
\\
 \label{predom}
\dot{\Gamma}(t)& = & - q^*~ \dot{X}(t) -
\frac{3 \sqrt \pi}{2 \lambda^* t^{3/2}}\mbox{Im} \left(\frac{1}{
    \sqrt{D}} \right)
+ {\cal O}\left( \frac{1}{t^2} \right) ~ .
\end{eqnarray}
As explained below, the coefficients $v^*, k^*=q^*+i\lambda^*$, 
and $D$ are all
given explicitly in terms of the dispersion relation of the linearized
equation. We shall focus on determining how these exact asymptotic
relaxation formulas  emerge, and why they are  independent
of the nonlinearities, the precise initial conditions, or on whether
the front dynamics is  regular
or chaotic. Before embarking on this, however,  
it is important to explain what we mean by velocity and 
phase for the various types of fronts.

{\em Uniformly translating pulled fronts}. The simplest types of fronts are
those for which the dynamical field 
$\phi(x,t)$ asymptotically approaches a uniformly translating profile
$\phi\equiv \Phi_{v^*}(\xi)$, $\xi\!=\!x\!-\!v^*t$, as happens, e.g., in the
celebrated nonlinear diffusion 
equation $\partial_t \phi=\partial_x^2 \phi + \phi-\phi^3 $ for fronts 
propagating into the unstable $\phi=0$ state. If
we define {\em level curves} as the lines in an $x,t$ diagram  where
$\phi(x,t)$ has a particular value, we can define the
velocity $v(t)$ as the slope of a level curve. For {\em uniformly
  translating fronts},  $q^*\!=\!0\!=\!\mbox{Im}D$;
(\ref{predv2}) then reduces to the expression derived for uniformly
translating fronts   in
\cite{evs}. The remarkable point is that the expression for $v(t)$ is
in this case completely independent of which level curve one 
 traces.
Moreover, it was shown in \cite{evs} that the nonlinear front region 
is slaved to the leading edge of the front whose  velocity relaxes 
according to (\ref{predv2}). This results in
\begin{eqnarray}
\label{uniform}
\phi(x,t) & = & \Phi_{v(t)}(\xi_X) +{\cal O}(t^{-2})~,~~\xi_X \ll \sqrt{t}~, 
\\ \xi_X & = & x-v^*t-X(t)~,\label{xix}
\end{eqnarray}
where $\Phi_{v}(\xi)$, $\xi\!=\!x\!-\!vt$ solves the {\em o.d.e.}\
for a front propagating uniformly with velocity $v$. $v(t)$ in 
(\ref{uniform}) is the instantaneous velocity of the front, and the 
frame $\xi_X$ is shifted by the time dependent quantity $X(t)$. 
Since the collective coordinate $X(t)$ diverges as $  \ln t$ for large
$t$ according to (\ref{predv2}), the difference between
$\xi_X$ and a uniformly translating frame is crucial --- only in the
former can we follow the relaxation.  Uniformly translating fronts have
no phase, hence all terms in (\ref{predom}) vanish identically.
 
{\em Coherent pattern generating fronts.} As an example of coherent
pattern generating fronts, we consider the so-called Swift-Hohenberg (SH)
equation
\begin{equation}\label{sh}
\partial_t u =\varepsilon u-(1+\partial_x^2)^2u -u^3
~~,~\varepsilon>0~.
\end{equation}

The space-time plot of Fig.\ 1{\em (a)} illustrates how SH-fronts 
with steep initial conditions (falling off faster than $e^{-\lambda^*x}$ 
as $x\! \to\! \infty$ into the unstable state $u\!=\!0$) generate a
periodic pattern. 
It is known that they are pulled \cite{vs2,dee,spruijt}.
In this case, new level curves in an $x,t$ plot are constantly 
being generated. If we define in this case the
velocity as the slope of the uppermost level curve, one gets an 
oscillatory function. Its average is $v(t)$ given
in (\ref{predv}), but $v(t)$ is difficult to extract this
way. Numerically, it is better to determine the velocity from an
empirical envelope obtained by interpolating the positions of the
maxima. Since these pattern forming front solutions for long times
have a temporal periodicity $u(\xi,t)= u(\xi,t+T)$ in the frame 
$\xi = x-vt$ moving with the velocity $v$ of the front, the 
asymptotic profiles can be written in the form $\sum_{n=1} e^{-2\pi i n t/T}
U_v^n(\xi)+ c.c.$. In terms of these complex modes $U$, our result for the
shape relaxation of the pulled front profile becomes in analogy to
(\ref{uniform})
\begin{equation}
u(x,t) \simeq \sum_{n=1} e^{- n i \Omega^* t - n i \Gamma(t)} U^n_{v(t)}
(\xi_X)+c.c. + \cdots \label{shasymp}
\end{equation}
with the frequency $\Omega^*$ given below \cite{remark3}.
Eq.\ (\ref{shasymp}) shows that  $\Gamma(t)$ is the {\em global phase} 
of the relaxing profile, as the functions $U_v^n$ only have a 
$\xi_X$-dependence. The result of our calculation of the long time 
relaxation of $v(t)$ and $\Gamma(t)$ is given in (\ref{predv}) --
(\ref{predom}). We stress that while for $\varepsilon \to 0$,
an ansatz like (\ref{shasymp}) leads to an amplitude equation
for the $n=\pm 1$ terms, our analysis applies for {\em any} $\varepsilon>0$.

{\em Incoherent or chaotic fronts.} The third class we
consider consists of fronts which leave behind chaotic states. They
occur in some regions of parameter space in the cubic Complex 
Ginzburg-Landau equation \cite{NB} or
in the quintic extension (QCGL) \cite{vsh} that we consider here,
\begin{eqnarray} \label{qcgle}
\partial_t A  =  \varepsilon A+(1+iC_1)\partial_x^2A&+&(1+iC_3)|A|^2
A  \nonumber \\ 
& - & (1-iC_5)|A|^4 A ~.
\end{eqnarray}
Fig.\ 1{\em (b)} shows an example of a pulled front in this
equation. Level curves in a space-time diagram
can now also both start and end. If we calculate the velocity 
from the slope of the uppermost level line, then its average value 
is again given by (\ref{predv2}) \cite{remark5}, but the oscillations 
can be quite large. However, our analysis confirms what is already visible  
in Fig.\ 1{\em (b)}, namely that even a chaotic pulled front 
becomes  more coherent the further one looks into 
the leading edge of the profile. Indeed we will see that 
in the leading edge where $|A| \ll 1$ the 
profile is given by an expression reminiscent of (\ref{shasymp}),
\begin{equation}
A(x,t) \approx  e^{-i\Omega^*
  t -i\Gamma(t)} e^{i k^* \xi_X}\psi(\xi_X), ~~{1 \ll\xi_X \ll \sqrt{t} }. 
  \label{cglasym}
\end{equation}
The fluctuations about this expression become smaller the larger $\xi_X$.

In Figs.\ 1{\em (c)} and 2{\em (c)} we show as an example results 
of our simulations of the SH equation (\ref{sh}) and the QCGL 
(\ref{qcgle}). They fully confirm our predictions (\ref{predv2}) 
and (\ref{predom}) for the asymptotic average velocity and phase 
relaxation. Note that for the QCGL, the fluctuations are indeed 
smaller the more one probes the leading edge region.

We now summarize how these results arise.

{\em Calculation of the asymptotic parameters.} Although this is
well-known \cite{ll}, we first briefly summarize how the linear
spreading velocity $v^*$ and the associated parameters $\lambda^*$
etc. arise, as the analysis also motivates the subsequent steps. After 
linearization about the unstable state, the equations we consider 
can all be written in the
form $\partial_t \phi= {\cal L}(\partial_x,\partial_x^2,\cdots) \phi$. For 
a Fourier mode $e^{-i\omega t +ikx}$, this yields the dispersion
relation $\omega(k)$. The linear spreading velocity $v^*$ of 
steep
initial conditions is then obtained by a saddle point analysis of the
Green's function $G$ of these equations. In the asymptotic frame
$\xi=x-v^*t$, $G(\xi,t)$ becomes
\begin{equation}
G(\xi,t)= \int \frac{{\rm d}k}{2\pi} e^{-i\Omega(k) t +ik\xi} 
 \approx
e^{ik^* \xi-i\Omega^*t}\frac{e^{-\frac{\xi^2}{4Dt}}}{\sqrt{4\pi Dt}}
 \label{greens}
\end{equation} for large times. Here $\Omega(k)=\omega(k)-v^*k$, and
\begin{equation}\label{lms}
\left. \frac{{\rm d} \Omega(k)}{{\rm d} k} \right|_{k^*}\!\!=\!0~, ~~~ 
\mbox{Im}\,\Omega(k^*)\!=\!0~,~~ ~  D\!=\! \left. \frac{i {\rm d}^2
  \Omega(k)}{2 {\rm d} k^2} \right|_{k^*}.
\end{equation}
The first equation in (\ref{lms}) is the saddle point condition,
while the second one expresses the self-consistency condition that
there is no growth in the co-moving frame. These equations straight
forwardly determine $v^*, k^*=q^*+i\lambda^*,D$ and the real frequency 
$\Omega^*=\Omega(k^*)$ \cite{results}.

{\em Choosing the proper frame and transformation.}
Eq.\ (\ref{greens}) not only confirms that a localized initial condition will
grow out and spread  asymptotically with the velocity $v^*$ given by
(\ref{lms}), but the Gaussian factor  also determines how the
asymptotic velocity is approached in the fully linear case. Our aim now is  to
understand the convergence of a pulled front
due to the interplay of the linear spreading and the
nonlinearities. The Green's function expression (\ref{greens}) gives
three important hints in this regard: 
First of all, $G(\xi,t)$ is
asymptotically of the form $e^{ik^*\xi -i \Omega^* t}$ {\em times} a
crossover function whose diffusive behavior is betrayed by the
Gaussian form in (\ref{greens}). Hence if we write our dynamical
fields as 
$A=e^{ik^*\xi -i \Omega^* t} \psi(\xi,t)$ for the QCGL (\ref{qcgle}) or 
$u=e^{ik^*\xi -i \Omega^* t} \psi(\xi,t)+ c.c.$ for the real field $u$
in (\ref{sh}), we expect that the dynamical equation for
$\psi(\xi,t)$ obeys a diffusion-type equation. Second, 
as we have argued in \cite{evs}, for the relaxation analysis one wants 
to work in a frame where the crossover function $\psi$ becomes 
asymptotically time independent. This is obviously not
true in the $\xi$ frame, due to the factor $1/\sqrt{t}$ in
(\ref{greens}). However, this term can be  absorbed
in the exponential prefactor, by writing 
$t^{-\nu}e^{ik^*\xi-i\Omega^*t}= e^{ik^*\xi-i\Omega^*t-\nu \ln t}$. 
Hence, we introduce the logarithmically shifted frame $ \xi_X\!=\!\xi\!-\!X(t)$
\cite{evs} as already used in (\ref{xix}). Third, we find a new feature
specific for pattern forming fronts: the complex parameters, and $D$ 
in particular, lead us to introduce the global phase $\Gamma(t)$.
We expand $\dot \Gamma(t)$ like $\dot X(t)$ \cite{evs}
\begin{equation} \label{xg}
\dot X(t)=\frac{c_1}{t}+\frac{c_{3/2}}{t^{3/2}}
+\cdots ~,~~~ \dot{\Gamma}(t)= \frac{d_1}{t}+\frac{d_{3/2}}{t^{3/2}}+\cdots
\end{equation} 
and analyze the long time dynamics by performing a ``leading edge
transformation'' to the field $\psi$,
\begin{eqnarray} \nonumber 
\mbox{QCGL:} ~~~~ A &=& e^{ik^*\xi_X-i\Omega^*t-i\Gamma(t)} \;\psi(\xi_X,t)~, 
\\ \label{shlet}
\mbox{SH:} ~~~~ u& = & e^{ik^*\xi_X-i\Omega^*t-i\Gamma(t)} \;\psi(\xi_X,t) 
+c.c.
\end{eqnarray}
Steep initial conditions imply that $\psi(\xi_X,t)\!\to\!0$ as 
$\xi_X\!\to\!\infty$. The determination of the coefficients 
in the expansions (\ref{xg}) of $\dot{X}$ and $\dot{\Gamma}$ is the main 
goal of the subsequent analysis, as this then directly yields Eqs.\
(\ref{predv2}) and (\ref{predom}). 

{\em Understanding the intermediate asymptotics.} Substituting 
the leading edge transformation (\ref{shlet}) into the nonlinear 
dynamical equations, we get 
\begin{eqnarray}
\partial_t\psi   & = &  D \partial_{\xi_X}^2\psi + \sum_{n=3} D_n
\partial_{\xi_X}^n \psi \nonumber \\ \label{xixeq}
&+& [ \dot{X} (t)(\partial_{\xi_X}+ik^*)+i\dot{ \Gamma}(t)] \psi - N(\psi) ~,
\end{eqnarray}
with $D_n= (-i/n!) {\rm d}^n \Omega/({\rm d}ik)^n|_{k^*}$ the generalization 
of $D$ in (\ref{lms}) (of course, for the QCGL, $\Omega(k)$ is quadratic 
in $k$, so $D_n=0$). In this equation, $N$ accounts for the nonlinear terms; 
e.g., for the QCGL, we simply have 
\begin{equation} \label{Neq}
N=   e^{-2\lambda^* \xi_X}|\psi|^2 \psi \;[ 1\!-\! iC_3\!+\!(1\!-\!iC_5)
  e^{-2\lambda^* \xi_X}|\psi|^2]~.
\end{equation}
The expression for the SH equation is similar. 

The structure of Eq.\ (\ref{xixeq}) is that of a diffusion-type
equation with $1/t$ and higher order
corrections from the $\dot{X}$ and $\dot{\Gamma}$
terms, and with a nonlinearity $N$. The crucial point to recognize now 
is that for fronts, $N$ is nonzero {\em only in a region of finite
  width}: For $\xi_X\!\to\!\infty$, 
  $N$ decays exponentially due to the explicit 
  exponential factors in (\ref{Neq}). 
For $\xi_X\! \to\!-\infty$, $N$ also decays exponentially, since
 $u$ and $A$ remain finite, so that $\psi$ decays as $e^{-\lambda^*|\xi_X |}$
 according to (\ref{shlet}). Intuitively, therefore,
we can think of  (\ref{xixeq}) as a diffusion 
equation in the presence of a sink $N$ localized 
at some finite value of $\xi_X$. The ensuing dynamics of $\psi$ 
to the right of the sink can be understood with the aid of 
Figs.\ 2{\em (a)} and {\em (b)}, which are obtained 
directly from the time-dependent numerical
simulations of the QCGL (\ref{qcgle}). To extract the intermediate
asymptotic behavior illustrated by these plots, we integrate
(\ref{xixeq}) once to get
\begin{eqnarray}
\label{inter}
& \partial_t & \int_{-\infty}^{\xi_X} \!\! {\rm d}\xi'_X \,\psi
=D\partial_{\xi_X}\psi + \sum_{n=3} \frac{D_n}{n-1}\;
\partial^{n-1}_{\xi_X} \psi  +  \\
&  & + i[k^* \dot X(t)+ \dot{\Gamma}(t)] \int_{-\infty}^{\xi_X} \!\! {\rm
  d}\xi'_X \, \psi  +  \dot X(t)\psi 
- \int_{-\infty}^{\xi_X} \!\! {\rm d}\xi'_X \, N(\psi)  
\nonumber
\end{eqnarray}
Now, in the region labeled I in Fig.\ 2{\em (b)}, we have for fixed
$\xi_X$ and $t \rightarrow \infty$ that the terms proportional to
$\dot{X}$ and $\dot{\Gamma}$ can be neglected upon averaging over 
the fast fluctuations; the same holds for the term on
the left. Since the integral converges quickly to the right due to the
exponential factors in $N$, we then get immediately, irrespective 
of the presence of higher order spatial derivatives
\begin{equation}
\lim_{t\rightarrow \infty} D \overline{{\frac{\partial \psi}{\partial
\xi_X}}} = \int_{-\infty}^{\infty} d\xi_X \overline{N(\psi)} \equiv \alpha D~.
\end{equation}
Here, the overbar denotes a time average (necessary for the case of a
chaotic front). Thus, for large times in region I, $\overline{\psi} \approx
\alpha \xi_X + \beta$ in dominant order. Moreover, from the diffusive nature
of the equation, our assertion that  the
fluctuations of $\psi$ rapidly decrease to the right of the region
where  $N$ is nonzero comes out naturally. In other words, provided
that the time-averaged sink 
strength $\alpha$ is nonzero, $\alpha \neq 0$, one will
find a buildup of a linear gradient in $|\overline{\psi}|$ in region I, {\em
  independent of the precise form of the nonlinearities or of whether
  or not the front dynamics is coherent}. This behavior is clearly
visible in Fig.\ 2{\em (b)}. We can  understand the dynamics in regions II 
and III along similar lines. In region III the
dominant terms in (\ref{xixeq}) are the one
on the left and the first one on the second line, and the cross-over
region II which separates regions I and III moves to the right
according to the diffusive law $\xi_X \sim D
\sqrt{t} $.

{\em Systematic expansion.}
These considerations are fully corroborated by our extension of 
the analysis of \cite{evs}. Anticipating that $\psi$ falls off  
for $\xi_X\gg1$, we split off a Gaussian factor by writing 
$\psi(\xi_X,t)=G(z,t)\;e^{-z} $ in terms of the similarity variable $
z={\xi_X^*}^2/(4Dt)$, 
 and expand
\begin{equation}
G(z,t)=t^{1/2}g_{-\frac{1}{2}}(z)+g_0(z)+t^{-1/2}g_{\frac{1}{2}}(z)+ \cdots
\end{equation} This, together with the expansion (\ref{xg}) for $X(t)$
and 
$\Gamma(t)$, the left ``boundary condition'' that $\psi(\xi_X,
t\! \rightarrow \! \infty) = \alpha \xi_X + \beta$  and the condition that
the functions $g(z)$ do not diverge exponentially, then results in  the
expressions (\ref{predv2}) for $\dot{X}(t)$ and (\ref{predom}) for
$\dot{\Gamma}$ \cite{spruijt}. For the QCGL, the analysis immediately
implies the result (\ref{cglasym}) for the front
profile in the leading edge. In addition for the SH equation, one
arrives at (\ref{shasymp}) for the shape relaxation in the front
interior along the lines of \cite{evs}: Starting from the {\em
  o.d.e.}'s for the $U^n_v$, one finds upon transforming to the frame
$\xi_X$ that to ${\cal O}(t^{-2})$, the time dependence only enters
parametrically through  $v(t)$. This then yields (\ref{shasymp}). 

In conclusion, we have shown that the long time relaxation of pulled
fronts is remarkably universal: Independent of whether fronts are
uniformly translating, pattern generating or chaotic, the velocity and
phase relaxation is governed by one simple formula, with universal
dominant and subdominant power law expressions.

\end{multicols}

\vspace{-.4cm}

\begin{figure}
\begin{center} 
\epsfig{figure=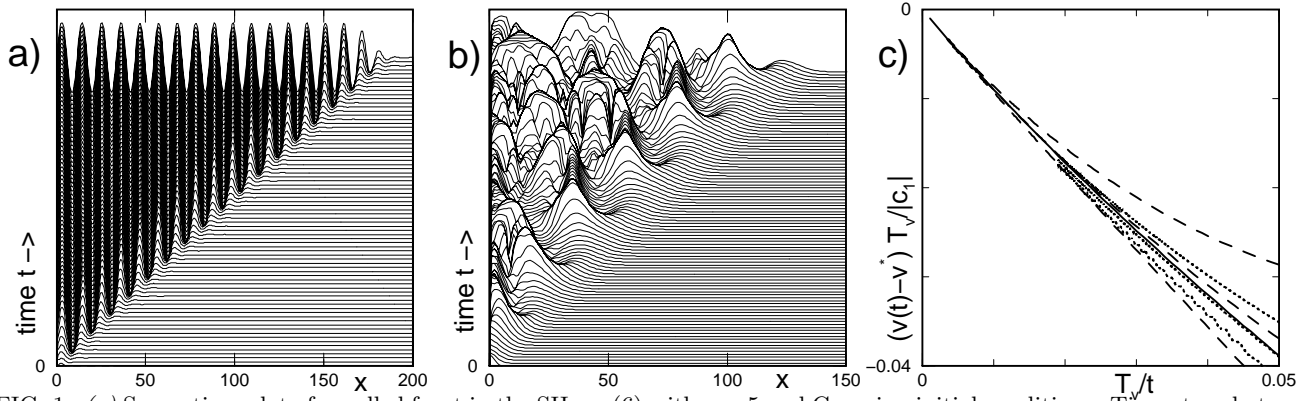}               
\caption[]{{\em (a)} Space-time plot of a pulled front 
in the SH eq.\ (\ref{sh}) with
  $\varepsilon= 5$ and Gaussian initial conditions. 
  Time steps between successive lines are 0.1.
  {\em (b)} A pulled
  front in the QCGL eq.\ (\ref{qcgle}) with
  $\varepsilon=0.25$, $C_1=1$, $C_3=C_5=-3$, and Gaussian 
  initial conditions. Plotted is $|A(x,t)|$.
  Time steps between lines are 1.
  {\em (c)} Scaling plot of the velocity relaxation
  $(v(t)-v^*)\cdot T_v/|c_1|$
  vs.\ $1/\tau$ with $\tau=t/T_v$ and characteristic time 
  $T_v=(c_{3/2}/c_1)^2$.
  Plotted are from top to bottom the data for the SH eq.\ for heights
  $u=0.0001\sqrt{\epsilon}$, $0.01 \sqrt{\epsilon}$, and
  $\sqrt{\epsilon}$ ($\epsilon=5$) as dashed lines, 
  and for the QCGL eq.\ (\ref{qcgle}) for heights
  $|A|=0.00002$, 0.0002, and 0.002 as dotted lines.
  The solid line is the universal asymptote
  $-1/\tau+1/\tau^{3/2}$.
}
\label{plots1}
\end{center}
\end{figure}

\vspace{-1.cm}

\begin{figure}
\begin{center}
\epsfig{figure=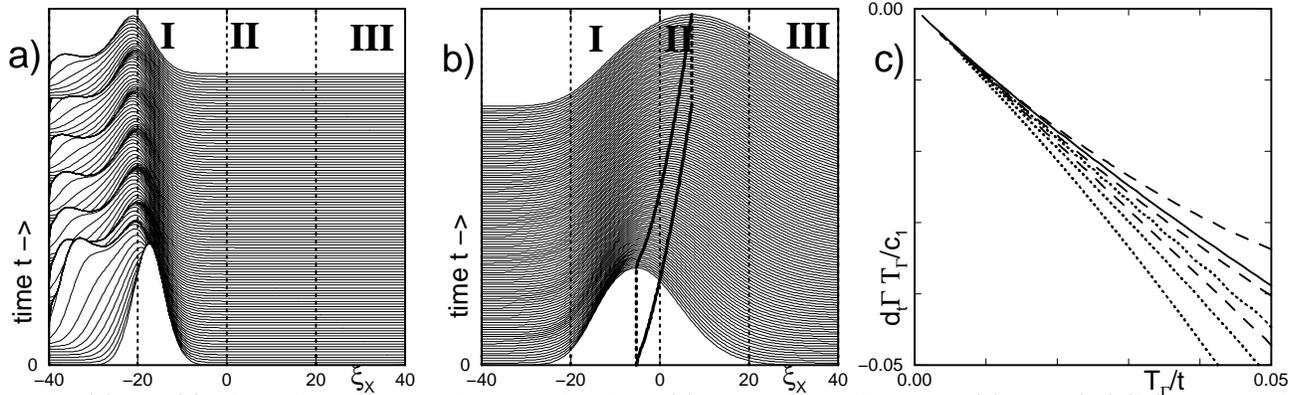}
\caption[]{{\em (a)} and {\em (b)}: Simulation of the QCGL eq.\
as in Fig.\ 1 {\em (b)} for times $t=35$ to 144. 
{\em (a)} shows $|N|$ (\ref{Neq}) as a function of $\xi_X$. 
{\em (b)} shows $|\psi|$, which in region I builds up a linear 
slope $\psi\propto\alpha\xi_X$, and in region III decays l
ike a Gaussian widening in time. The lines in region II show 
the maxima of $\psi(\xi_X,t)$ for fixed $t$ and their projection 
$\xi_X\sim\sqrt{t}$ into the $(\xi_X,t)$ plane.
{\em (c)} shows the scaling plot for the phase relaxation.
From top to bottom: SH (dashed) for $u=0.0001\sqrt{\epsilon}$, 
$0.01 \sqrt{\epsilon}$, and $\sqrt{\epsilon}$ ($\epsilon=5$),
and QCGL (dotted) for $|A|=0.002$, 0.0002, and 0.00002.
Plotted is $\dot\Gamma(t) \cdot T_\Gamma/c_1$ 
vs.\ $1/\tau$. Here $\tau=t/T_\Gamma$, and 
$T_\Gamma=T_v\cdot\left[1+\lambda^*\mbox{Im}D^{-1/2}/
(q^*\mbox{Re}D^{-1/2})\right]$. The solid line again is 
the universal asymptote $-1/\tau+1/\tau^{3/2}$.
}
\label{plots2}
\end{center}
\end{figure}

\begin{multicols}{2}

\end{multicols}

\begin{references}
\bibitem{CH} M. C. Cross and P. C. Hohenberg, Rev. Mod. Phys. {\bf
    65}, 851 (1993). 
\bibitem{bio}  J.D. Murray, {\em Mathematical Biology} (Springer,
  Berlin, 1989). 
\bibitem{bj} E.\ Ben-Jacob, H.R.\ Brand, G.\ Dee, L.\ Kramer, and
  J.S.\ Langer, Physica D {\bf 14}, 348 (1985).
\bibitem{vs2} W. van Saarloos, Phys. Rev. A {\bf 39}, 6367 (1989).
\bibitem{stokes} A. N. Stokes, Math. Biosci. {\bf 31},  307 (1981).
\bibitem{paq} G. C. Paquette, L.-Y. Chen, N. Goldenfeld and  Y. Oono,
    Phys. Rev. Lett. {\bf72}, 76 (1994).
\bibitem{evs} U.\ Ebert and W.\ van Saarloos, Phys.\ Rev.\ Lett.\
  {\bf 80}, 1650 (1998); preprint submitted to Physica D (available at
{\scriptsize{\tt  http://www.cwi.nl/static/publications/reports/MAS-1999.html}}).
\bibitem{dee} G. Dee and J. S. Langer, Phys. Rev. Lett. {\bf 50}, 383
  (1983).
\bibitem{spruijt} U. Ebert, W. Spruijt and W. van Saarloos
  (unpublished).
\bibitem{remark3} Underlying  (\ref{shasymp}) is the assumption that
  the (\ref{sh}) admits a two-parameter family of front
  solutions. This was shown  for small
  $\varepsilon$ by P. Collet and J.-P. Eckmann, Commun. Math. Phys. 
  {\bf 107}, 39 (1986),  and is demonstrated by counting arguments for
  arbitrary $\varepsilon$ in \cite{spruijt}.
\bibitem{NB} K. Nozaki and N. Bekki, Phys. Rev. Lett. {\bf 51}, 271
  (1983).
\bibitem{vsh} W. van Saarloos and P. C. Hohenberg, Physica D {\bf 56},
303  (1992).
\bibitem{remark5} This is true for chaotic fronts provided that the
  temporal correlation function for the chaotic variable 
  falls off at least as fast as $t^{-2}$, so that the temporal
  change of the average velocity $v(t)$ can be considered adiabatically.
\bibitem{ll} E. M. Lifshitz and L.P. Pitaevskii, {\em Physical Kinetics}
  (Pergamon, New York, 1981).
\bibitem{results} For (\ref{qcgle}), $v^*=2\sqrt{\varepsilon (1+C_1^2)}, \ 
k^*=(C_1+i)\sqrt{\varepsilon /(1+C_1^2)}$,
$\Omega^*=-C_1\varepsilon$ and $D=(1+iC_1)$. For (\ref{sh}),
$\lambda^*=[(\sqrt{1+6\varepsilon}-1)/12]^{1/2}$, $q^*=\pm
\sqrt{1+3\lambda^{*2}}$, $v^*=8\lambda^*(1+4\lambda^{*2})$,
$\Omega^*=-8\lambda^*q^{*3}$, $D=4 q^{*2}+12iq^*\lambda^*$.
\end{references}
\end{document}